\documentclass[11pt]{article}
\usepackage{amssymb,amsmath,amsfonts}
\usepackage{graphicx}
\usepackage{graphics}
\usepackage{eepic,epsfig}

\begin{document}

\title{Synchrotron radiation inside a dielectric cylinder}
\author{A. A. Saharian$^{1,2}$\thanks{%
Email address: saharian@ysu.am}, \thinspace\ A. S. Kotanjyan$^{1}$ \\
\vspace{0.2cm}\\
\textit{$^{1}$Department of Physics, Yerevan State University,}\\
\textit{1 Alex Manoogian Street, 0025 Yerevan, Armenia}\\
\vspace{0.2cm}\\
\textit{$^{2}$Institute of Applied Problems in Physics,}\\
\textit{National Academy of Sciences of the Republic of Armenia,}\\
\textit{25 Nersisyan Street, 0014 Yerevan, Armenia}}
\date{\today}
\maketitle

\begin{abstract}
We investigate the electromagnetic fields generated by a charged particle
rotating inside a dielectric cylinder immersed into a homogeneous medium.
The expressions for the bound modes of the radiation field are derived in
both interior and exterior regions. The radiation intensity for the modes
propagating inside the cylinder is evaluated by using two different ways: by
evaluating the work done by the radiation field on the charge and by
evaluating the energy flux through the cross-section of the cylinder. The
relation between these two quantities is discussed. We investigate the
relative contributions of the bound modes and the modes propagating at large
distances from the cylinder to the total radiation intensity. Numerical
examples are given for a dielectric cylinder in the vacuum. It is shown that
the presence of the cylinder can lead to the considerable increase of the
synchrotron radiation intensity.
\end{abstract}

\bigskip

PACS number(s): 41.60.Ap, 41.60.Bq

\bigskip

\section{Introduction}

\label{sec:oscint}

Synchrotron radiation is an important tool in many disciplines. Its unique
characteristics, such as high intensity and high collimation, have resulted
in extensive applications (see, for example, \cite{Soko86,Bord99,Hofm04}).
Motivated by this, the investigation of mechanisms for the control of the
radiation parameters is of great interest. Particularly important is the
study of the influence of medium on the spectral and angular characteristics
of the synchrotron radiation. The presence of medium can essentially change
the characteristics of the high-energy electromagnetic processes. Moreover,
new types of phenomena arise. Well-known examples are Cherenkov and
transition radiations.

In \cite{Grig95}-\cite{Saha05} we have investigated the synchrotron
radiation from a charge rotating around/inside a dielectric cylinder
enclosed by a homogeneous medium. It has been shown that under the Cherenkov
condition for the material of the cylinder and the particle velocity, strong
narrow peaks appear in the angular distribution of the radiation intensity
in the exterior medium. At these peaks the radiated energy exceeds the
corresponding quantity in the case of a homogeneous medium by several orders
of magnitude. Similar features for the radiation generated by a charge
moving along a helical orbit have been discussed in \cite{Saha07a}-\cite%
{Saha09}. The case for an electron rotating around/inside a dielectric ball
has been considered in \cite{Grig95b}-\cite{Grig05}.

In the previous investigations of the radiation from a charge rotating
around/inside a dielectric cylinder we have considered the radiation
intensity in the exterior region at large distances from the cylinder. In
addition to this part there is radiation which propagates inside the
cylinder and it is of interest to investigate the relative contribution of
these modes to the total intensity. In the present paper we consider the
radiation intensity inside a dielectric cylinder emitted by a charge
rotating inside the cylinder.

The paper is organized as follows. In the next section, we present the
formulas for the magnetic and electric fields inside a dielectric cylinder.
The parts corresponding to the radiation fields are separated in Section \ref%
{sec:RadFields}. These parts are emitted on the eigenmodes of the cylinder.
The radiation intensity inside the dielectric cylinder is considered in
Section \ref{sec:IntensInt}. Two quantities are evaluated: the work done by
the radiation field on the charge and the energy flux passing through the
cross-section of the cylinder. We show that these two quantities differ. In
Section \ref{sec:IntExt} we investigate the part of the radiation intensity
in the exterior region corresponding to the eigenmodes of the dielectric
cylinder and exponentially decreasing outside the cylinder. It is shown that
the difference between the intensity evaluated by the work done by the
radiation field and the intensity evaluated by the flux through the
cross-section of the cylinder is due to the energy flux carrying by the
eigenmodes of the cylinder in the exterior region. The main results are
summarized in Section \ref{sec:Conc}.

\section{Electromagnetic fields inside a dielectric cylinder}

\label{sec:Fields}

We consider a dielectric cylinder of radius $\rho _{1}$ and with dielectric
permittivity $\varepsilon _{0}$ and a point charge $q$ rotating inside a
cylinder. The radius of the rotation orbit and the velocity of the charge
will be denoted by $\rho _{0}$, $\rho _{0}<\rho _{1}$, and $v$ respectively.
We assume that the system is immersed in a homogeneous medium with
permittivity $\varepsilon _{1}$. In a properly chosen cylindrical coordinate
system $(\rho ,\phi ,z)$ with the $z$-axis directed along the cylinder axis,
the components of the current density created by the charge are given by the
formula
\begin{equation}
j_{\phi }=\frac{q}{\rho }v\delta (\rho -\rho _{0})\delta (\phi -\omega
_{0}t)\delta (z),  \label{hosqxtutjun}
\end{equation}%
where $\omega _{0}=v/\rho _{0}$ is the angular velocity of the charge. In
accordance with the problem symmetry, the electric and magnetic fields can
be presented in the form of the Fourier expansion%
\begin{equation}
F_{l}(\mathbf{r},t)=\sum_{m=-\infty }^{\infty }e^{im(\phi -\omega
_{0}t)}\int_{-\infty }^{\infty }dk_{z}e^{ik_{z}z}F_{ml}(k_{z},\rho ),
\label{FlFour}
\end{equation}%
where $F=E,H$ for electric and magnetic fields respectively. As the
functions $F_{l}(\mathbf{r},t)$\ are real, one has $F_{ml}^{\ast
}(k_{z},\rho )=F_{-ml}(-k_{z},\rho )$. Consequently, formula (\ref{FlFour})
can also be rewritten in the form
\begin{equation}
F_{l}(\mathbf{r},t)=2{\mathrm{Re}}\left[ \sideset{}{'}{\sum}_{m=0}^{\infty
}e^{im(\phi -\omega _{0}t)}\int_{-\infty }^{\infty
}dk_{z}e^{ik_{z}z}F_{ml}(k_{z},\rho )\right] ,  \label{vecpot21}
\end{equation}%
where the prime means that the term $m=0$ should be taken with the weight
1/2. In the discussion below we will assume that $m\geqslant 0$.

In the region $\rho _{0}<\rho <\rho _{1}$ the Fourier components of the
magnetic field, $H_{ml}=H_{ml}(k_{z},\rho )$, can be presented in the form
\cite{Kota02a}%
\begin{eqnarray}
H_{ml} &=&\frac{qvk_{z}}{4ci^{\sigma _{l}-1}}\sum_{p=\pm 1}p^{\sigma _{l}-1}%
\left[ J_{m+p}(\lambda _{0}\rho _{0})H_{m+p}(\lambda _{0}\rho
)+B_{1,m}^{(p)}J_{m+p}(\lambda _{0}\rho )\right] ,\;l=\rho ,\phi ,  \notag \\
H_{mz} &=&\frac{iqv\lambda _{0}}{4c}\sum_{p=\pm 1}p\left[ J_{m+p}(\lambda
_{0}\rho _{0})H_{m}(\lambda _{0}\rho )+B_{1,m}^{(p)}J_{m}(\lambda _{0}\rho )%
\right] ,  \label{Hmz}
\end{eqnarray}%
where $J_{m}(x)$ is the Bessel function, $H_{m}(x)=H_{m}^{(1)}(x)$ is the
Hankel function of the first kind, $\sigma _{\rho }=1$, $\sigma _{\phi }=2$,
and%
\begin{equation}
\lambda _{j}^{2}=m^{2}\omega _{0}^{2}\varepsilon _{j}/c^{2}-k_{z}^{2},\quad
j=0,1.  \label{lambj}
\end{equation}%
The coefficients $B_{1,m}^{(p)}$ are determined by the expressions
\begin{equation}
B_{1,m}^{(p)}=-J_{m+p}(\lambda _{0}\rho _{0})\frac{W_{m+p}^{H}}{W_{m+p}^{J}}+%
\frac{ip\lambda _{1}H_{m+p}(\lambda _{1}\rho _{1})}{\pi \rho _{1}\alpha
_{m}W_{m+p}^{J}}H_{m}(\lambda _{1}\rho _{1})\sum_{l=\pm 1}\frac{%
J_{m+l}(\lambda _{0}\rho _{0})}{W_{m+l}^{J}},  \label{Bm+p}
\end{equation}%
with the notations
\begin{equation}
\alpha _{m}=\frac{\varepsilon _{0}}{\varepsilon _{1}-\varepsilon _{0}}-\frac{%
1}{2}\lambda _{0}J_{m}(\lambda _{0}\rho _{1})\sum_{l=\pm 1}l\frac{%
H_{m+l}(\lambda _{1}\rho _{1})}{W_{m+l}^{J}},  \label{bet1}
\end{equation}%
and%
\begin{equation}
W_{m}^{F}=\lambda _{1}F_{m}(\lambda _{0}\rho _{1})H_{m}^{\prime }(\lambda
_{1}\rho _{1})-\lambda _{0}H_{m}(\lambda _{1}\rho _{1})F_{m}^{\prime
}(\lambda _{0}\rho _{1}),  \label{WronskianF}
\end{equation}%
with $F=J,H$. The corresponding expressions for the Fourier components of
the magnetic fields for $\rho <\rho _{0}$ are obtained from (\ref{Hmz}) by
the replacements $J\rightleftarrows H$ of the Bessel and Hankel functions in
the first terms in the square brackets of (\ref{Hmz}). The parts with these
terms correspond to the field of the charge in a homogeneous medium with
permittivity $\varepsilon _{0}$.

By making use of the Maxwell equation $\mathbf{E=}ic(\omega \varepsilon
_{0})^{-1}\nabla \times \mathbf{H}$, from (\ref{Hmz}) one can derive the
corresponding Fourier coefficients for the electric field, $%
E_{ml}=E_{ml}(k_{z},\rho )$, in the region $\rho _{0}<\rho <\rho _{1}$:
\begin{eqnarray}
E_{ml} &=&\frac{qvi^{-\sigma _{l}}}{8m\omega _{0}\varepsilon _{0}}%
\sum_{p=\pm 1}p^{\sigma _{l}}\left[ H_{m+p}(\lambda _{0}\rho )\sum_{j=\pm
1}\left( k_{z}^{2}+j\frac{m^{2}\omega _{0}^{2}\varepsilon _{0}}{c^{2}}%
\right) J_{m+jp}(\lambda _{0}\rho _{0})\right.  \notag \\
&&+\left. J_{m+p}(\lambda _{0}\rho )\sum_{j=\pm 1}\left( k_{z}^{2}+j\frac{%
m^{2}\omega _{0}^{2}\varepsilon _{0}}{c^{2}}\right) B_{1,m}^{(jp)}\right] ,
\notag \\
E_{mz} &=&\frac{qv\lambda _{0}k_{z}}{4m\omega _{0}\varepsilon _{0}}%
\sum_{p=\pm 1}\left[ J_{m+p}(\lambda _{0}\rho _{0})H_{m}(\lambda _{0}\rho
)+B_{1,m}^{(p)}J_{m}(\lambda _{0}\rho )\right] ,  \label{Emlz}
\end{eqnarray}%
where $l=\rho ,\phi $. The corresponding expressions in the region $\rho
<\rho _{0}$ are obtained from (\ref{Emlz}) by the replacements $%
J\rightleftarrows H$ in the first terms in the square brackets. The latter
correspond to the field of the charge in a homogeneous medium with
permittivity $\varepsilon _{0}$.

\section{Radiation fields inside a dielectric cylinder}

\label{sec:RadFields}

In this section we consider the radiation fields propagating inside a
dielectric cylinder. The radiation field is determined by the singular
points of the integrand in the integral over $k_{z}$ in (\ref{vecpot21}).
For the parts in the Fourier components corresponding to the fields in a
homogeneous medium with permittivity $\varepsilon _{0}$ the integrands are
regular and these parts do not contribute to the radiation field. In the
parts due to the inhomogeneity, the coefficients $B_{1,m}^{(p)}$ enter in
the form of the combinations $\sum_{p}B_{1,m}^{(p)}$ and $%
\sum_{p}pB_{1,m}^{(p)}$. By using formula (\ref{Bm+p}), it can be seen that
these combinations are regular at the points corresponding to the zeros of
the functions $W_{m}^{J}$ and $W_{m\pm 1}^{J}$. The only poles of the parts
of the Fourier components are the zeros of the function $\alpha _{m}$
appearing in the denominator of Eq. (\ref{Bm+p}). It can be shown that this
function has zeros only for $\lambda _{0}^{2}>0$ and $\lambda _{1}^{2}<0$.
In particular, there is no radiation on the mode $m=0$. Note that for the
corresponding modes in the exterior region (see Ref. \cite{Saha05}), $\rho
>\rho _{1}$, the Fourier coefficients are proportional to the MacDonald
function $K_{\nu }(|\lambda _{1}|\rho )$, with $\nu =m,m\pm 1$, and they are
exponentially damped in the region outside the cylinder. These modes are the
eigenmodes of the dielectric cylinder and propagate inside the cylinder. By
using the properties of the cylindrical functions, the function $\alpha _{m}$
can be rewritten in the form
\begin{equation}
\alpha _{m}=\frac{U_{m}}{(\varepsilon _{1}-\varepsilon _{0})\left(
V_{m}^{2}-m^{2}u^{2}\right) },  \label{alfmnew}
\end{equation}%
where we have used the notations%
\begin{eqnarray}
V_{m} &=&|\lambda _{1}|\rho _{1}\frac{J_{m}^{\prime }}{J_{m}}+\lambda
_{0}\rho _{1}\frac{K_{m}^{\prime }}{K_{m}},\;u=\frac{\lambda _{0}}{|\lambda
_{1}|}+\frac{|\lambda _{1}|}{\lambda _{0}},  \label{Vmrad} \\
U_{m} &=&V_{m}\left( \varepsilon _{0}|\lambda _{1}|\rho _{1}\frac{%
J_{m}^{\prime }}{J_{m}}+\varepsilon _{1}\lambda _{0}\rho _{1}\frac{%
K_{m}^{\prime }}{K_{m}}\right) -m^{2}\frac{\lambda _{0}^{2}+|\lambda
_{1}|^{2}}{\lambda _{0}^{2}|\lambda _{1}|^{2}}\left( \varepsilon _{1}\lambda
_{0}^{2}+\varepsilon _{0}|\lambda _{1}|^{2}\right) .  \label{Umrad}
\end{eqnarray}%
Here and below it is understood $K_{m}=K_{m}(|\lambda _{1}|\rho _{1})$, $%
J_{m}=J_{m}(\lambda _{0}\rho _{1})$ if the argument of the function is
omitted, and the prime means the differentiation with respect to the
argument of the function. Now the equation for the eigenmodes is written in
the standard form (see, for instance,~\cite{Jackson})%
\begin{equation}
U_{m}=0.  \label{eigmodesnew2}
\end{equation}%
For the eigenmodes of $k_{z}$ one has%
\begin{equation}
\frac{m^{2}\omega _{0}^{2}}{c^{2}}\varepsilon _{1}\leqslant
k_{z}^{2}\leqslant \frac{m^{2}\omega _{0}^{2}}{c^{2}}\varepsilon _{0}.
\label{kzCond}
\end{equation}%
Unlike to the case of the waveguide with perfectly conducting walls, here
there is no separation into purely transverse electric (TE) and transverse
magnetic (TM) modes.

We denote by $k_{z}=\pm k_{m,s}$, $k_{m,s}>0$, $s=1,2,\ldots ,s_{m}$, the
solutions to equation (\ref{eigmodesnew2}). The problem under consideration
is symmetric with respect to the replacement $z\rightarrow -z$ and in the
discussion below we consider the radiation in the region $z>0$ for which $%
k_{z}=k_{m,s}$. Instead of $k_{z}$ we can introduce an angular variable $%
\theta $ with the eigenvalues $\theta =\theta _{m,s}$ defined by%
\begin{equation}
\cos \theta _{m,s}=\frac{ck_{m,s}}{m\omega _{0}\sqrt{\varepsilon _{0}}}%
,\;\cos \theta _{m,s}\geqslant \sqrt{\varepsilon _{1}/\varepsilon _{0}},
\label{teta}
\end{equation}%
where the last relation follows from the left inequality in (\ref{kzCond}).
For fixed values of the other parameters the number of the modes $k_{m,s}$
increases with increasing $\rho _{1}$. In the discussion below we shall
consider the numerical examples for the electron energy $E_{e}=2$ MeV and
for $\varepsilon _{1}=1$, $\varepsilon _{0}=3$. For these values of the
parameters and for $\rho _{1}/\rho _{0}=1.2$ one has $s_{m}=1$ for $%
m=1,2,\ldots ,6$, $s_{m}=2$ for $m=7,8,9,10$, $s_{m}=3$ for $m=11,12,13,14$,
$s_{m}=4$ for $m=15,16$. In the table below we have presented the values of $%
\theta _{m,s}$ (in degrees) for $m=1-6$.

\begin{center}
\begin{tabular}{ccccccc}
\hline
$m$ & $1$ & $2$ & $3$ & $4$ & $5$ & $6$ \\ \hline
$\theta _{m,s}$ & $50.64$ & $51.15$ & $49.07$ & $46.84$ & $45.04$ & $43.61$
\\ \hline
\end{tabular}
\end{center}

The number of modes increases with increasing values of $\rho _{1}$. For
example, for $\rho _{1}/\rho _{0}=2$ and for $m=4$ we have $s_{m}=3$ with $%
\theta _{m,s}=$50.93, 43.77, 38.64 for $s=1,2,3$, respectively. For $\rho
_{1}/\rho _{0}=4$ and for $m=4$ we have $s_{m}=11$. For large values of $m$,
the equation for the eigenmodes (\ref{eigmodesnew2}) is simplified by using
the Debye's asymptotic expansions for the cylindrical functions (see, for
instance, \cite{Abra72}). From these expansions it follows that $%
K_{m}^{\prime }(mx)/K_{m}(mx)\sim -\sqrt{1+x^{2}}/x$. For the Bessel
function two separate cases should be considered. For $x<1$ one has $%
J_{m}^{\prime }(mx)/J_{m}(mx)\sim \sqrt{1-x^{2}}/x$ and for $x>1$ the
asymptotic takes the form $J_{m}^{\prime }(mx)/J_{m}(mx)\sim \tan \psi \sqrt{%
x^{2}-1}/x$, where $\psi $ is defined by the relations $\psi =m(\tan \beta
-\beta )-\pi /4$, $\tan \beta =\sqrt{x^{2}-1}$. By using these asymptotics
it can be seen that for large values of $m$, the equation for the eigenmodes
has no solutions when $\lambda _{0}\rho _{1}<m$.

In order to obtain unambiguous result for the fields given by (\ref{vecpot21}%
), we should specify the integration contour in the complex plane $k_{z}$.
For this we note that in physical situations the dielectric permittivity $%
\varepsilon _{0}$ is a complex quantity: $\varepsilon _{0}=\varepsilon
_{0}^{\prime }+i\varepsilon _{0}^{\prime \prime }$. Assuming that $%
\varepsilon _{0}^{\prime \prime }$ is small, this induces an imaginary part
for $k_{z}$ given by formula%
\begin{equation}
{\mathrm{Im\,}}k_{z}=\frac{im\omega _{0}}{2c}\varepsilon _{0}^{\prime \prime
}(m\omega _{0})\left[ \varepsilon _{0}^{\prime }(m\omega _{0})-\left( \frac{%
c\lambda _{m,s}}{m\omega _{0}\rho _{1}}\right) ^{2}\right] ^{-1/2}.
\label{Imkz}
\end{equation}%
Note that one has $\varepsilon _{0}^{\prime \prime }(m\omega _{0})>0$ and,
hence, in accordance with (\ref{Imkz}), ${\mathrm{Im\,}}k_{z}>0$. Deforming
the integration contour we see that in the integral over $k_{z}$ in (\ref%
{vecpot21}), the contour avoids the poles $k_{m,s}$ from below.

Now let us consider the electromagnetic fields inside the cylinder in the
region $z>0$ for large distances from the charge. In this case the dominant
contribution into the fields comes from the poles of the integrand. Having
specified the integration contour now we can evaluate the radiation parts of
the fields inside the cylinder. Closing the integration contour by the large
semicircle in the upper half-plane, one finds:%
\begin{equation}
F_{l}(\mathbf{r},t)=\frac{qv}{c}\sum_{m=1}^{\infty }\sum_{s=1}^{s_{m}}\frac{%
F_{m,s}^{l}(\rho )}{\alpha _{m}^{\prime }(k_{m,s})}R(m(\phi -\omega
_{0}t)+k_{m,s}{}z),  \label{Flr}
\end{equation}%
where $R(x)=\cos x$ for the components $H_{\rho }$, $E_{\phi }$, $E_{z}$,
and $R(x)=\sin x$ for the components $H_{\phi }$, $H_{z}$, $E_{\rho }$. In
the case of the magnetic field the functions for the separate components in (%
\ref{Flr}) are defined by%
\begin{eqnarray}
H_{m,s}^{l}(\rho ) &=&k_{m,s}\sum_{p=\pm 1}p^{\sigma
_{l}-1}B_{1,m,s}^{(p)}J_{m+p}(\lambda _{m,s}\rho /\rho _{1}),  \notag \\
H_{m,s}^{z}(\rho ) &=&-\frac{\lambda _{m,s}}{\rho _{1}}J_{m}(\lambda
_{m,s}\rho /\rho _{1})\sum_{p=\pm 1}pB_{1,m,s}^{(p)},  \label{Hrad}
\end{eqnarray}%
where%
\begin{equation*}
B_{1,m,s}^{(p)}=\frac{\lambda _{m,s}^{(1)}K_{m+p}/K_{m}}{\left(
V_{m}-pmu\right) J_{m}^{2}}\sum_{l=\pm 1}\frac{J_{m+l}(\lambda _{m,s}\rho
_{0}/\rho _{1})}{V_{m}-lmu},
\end{equation*}%
with $K_{m}=K_{m}(\lambda _{m,s}^{(1)})$, $J_{m}=J_{m}(\lambda _{m,s})$, and
we have introduced the notations
\begin{equation}
\lambda _{m,s}=\rho _{1}\sqrt{m^{2}\omega _{0}^{2}\varepsilon
_{0}/c^{2}-k_{m,s}^{2}},\;\lambda _{m,s}^{(1)}=\rho _{1}\sqrt{%
k_{m,s}^{2}-\varepsilon _{1}m^{2}\omega _{0}^{2}/c^{2}}.  \label{lams}
\end{equation}%
In terms of $\theta _{m,s}$ one has%
\begin{equation}
\lambda _{m,s}=\frac{m\omega _{0}\rho _{1}}{c}\sqrt{\varepsilon _{0}}\sin
\theta _{m,s},\;\lambda _{m,s}^{(1)}=\frac{m\omega _{0}\rho _{1}}{c}\sqrt{%
\varepsilon _{0}\cos ^{2}\theta _{m,s}-\varepsilon _{1}}.  \label{lams1}
\end{equation}

For the corresponding functions in the formulas of the components for the
electric field we find the expressions%
\begin{eqnarray}
E_{m,s}^{l}(\rho ) &=&\frac{(-1)^{\sigma _{l}-1}c}{2\varepsilon _{0}m\omega
_{0}}\sum_{p=\pm 1}p^{\sigma _{l}}J_{m+p}(\lambda _{m,s}\rho /\rho
_{1})\sum_{j=\pm 1}j\left( \frac{m^{2}\omega _{0}^{2}}{c^{2}}\varepsilon
_{0}+jk_{m,s}^{2}\right) B_{1,m,s}^{(jp)},  \notag \\
E_{m,s}^{z}(\rho ) &=&-\frac{c\lambda _{m,s}k_{m,s}}{\varepsilon _{0}m\omega
_{0}\rho _{1}}\sum_{p=\pm 1}B_{1,m,s}^{(p)}J_{m}(\lambda _{m,s}\rho /\rho
_{1}).  \label{Erad}
\end{eqnarray}%
Note that on the cylinder axis, $\rho =0$, one has $E_{z}(\mathbf{r}%
,t)=H_{z}(\mathbf{r},t)=0$.

\section{Radiation intensity inside a dielectric cylinder}

\label{sec:IntensInt}

Having the radiation fields we can evaluate the radiation intensity
propagating inside the dielectric cylinder. This can be done in two
different ways. In the first one we evaluate the work done by the radiation
field on the charged particle:%
\begin{equation}
I=-\int d\rho d\varphi dz\rho \,j_{\phi }E_{\phi }.  \label{Iwork1}
\end{equation}%
By taking into account the expressions for the current density (\ref%
{hosqxtutjun}) and for $E_{\phi }$ from (\ref{Erad}), for the radiation
intensity one finds%
\begin{eqnarray}
I &=&\sum_{m=1}^{\infty }I_{m}=\frac{q^{2}v^{2}}{2\varepsilon _{0}\omega _{0}%
}\sum_{m=1}^{\infty }\sum_{s=1}^{s_{m}}\frac{\lambda _{m,s}^{(1)}J_{m}^{-2}}{%
m\alpha _{m}^{\prime }(k_{m,s})K_{m}}\sum_{l=\pm 1}\frac{J_{m+l}(\lambda
_{m,s}\rho _{0}/\rho _{1})}{V_{m}-lmu}  \notag \\
&&\times \sum_{p=\pm 1}J_{m+p}(\lambda _{m,s}\rho _{0}/\rho _{1})\left[
\left( \frac{m^{2}\omega _{0}^{2}}{c^{2}}\varepsilon _{0}+k_{m,s}^{2}\right)
\frac{K_{m+p}}{V_{m}-pmu}-\frac{\lambda _{m,s}^{2}}{\rho _{1}^{2}}\frac{%
K_{m-p}}{V_{m}+pmu}\right] .  \label{Iwork2}
\end{eqnarray}%
Note that in this formula we have $\alpha _{m}^{\prime
}(k_{m,s})=U_{m}^{\prime }(k_{m,s})\left[ (\varepsilon _{1}-\varepsilon
_{0})\left( V_{m}^{2}-m^{2}u^{2}\right) \right] ^{-1}$.

Let us consider the limiting case of formula (\ref{Iwork2}) when $\rho
_{1}\rightarrow \infty $. From the equation (\ref{eigmodesnew2}) it follows
that in this limit we have two types of the modes. For the first one
\begin{equation}
|\lambda _{1}|\frac{J_{m}^{\prime }}{J_{m}}+\lambda _{0}\frac{K_{m}^{\prime }%
}{K_{m}}\approx -|\lambda _{1}|\tan \left( \lambda _{0}\rho _{1}-\frac{m}{2}%
\pi -\frac{\pi }{4}\right) -\lambda _{0}=0,  \label{limMode1}
\end{equation}%
where we have used the asymptotic formulas for the Bessel functions for
large values of the arguments. For this type of modes, the expression (\ref%
{Iwork2}) for the radiation intensity takes the form%
\begin{equation}
I_{(1)}\approx \frac{2\pi q^{2}v^{2}}{\rho _{1}c^{2}}\sum_{m=1}^{\infty
}\sum_{s}\frac{m\omega _{0}\lambda _{0}J_{m}^{\prime 2}(\lambda _{0}\rho
_{0})}{\sqrt{\left( m\omega _{0}/c\right) ^{2}\varepsilon _{0}-\lambda
_{0}^{2}}}.  \label{Ilim1}
\end{equation}%
The equation (\ref{limMode1}) for the eigenmodes can also be written in the
form%
\begin{equation}
\lambda _{0}\rho _{1}-\frac{m}{2}\pi -\frac{\pi }{4}+\gamma =\pi s,
\label{eigen1}
\end{equation}%
where $\gamma =\arcsin [c\lambda _{0}/(m\omega _{0}\sqrt{\varepsilon
_{0}-\varepsilon _{1}})]$. From here it follows that in the limit $\rho
_{1}\rightarrow \infty $ one has $\lambda _{0}\approx \pi s/\rho _{1}$. Now
we see that the dominant contribution to (\ref{Ilim1}) comes from large
values $s$ and we can replace the summation over $s$ by the integration: $%
\sum_{s}\rightarrow \int ds=(\rho _{1}/\pi )\int d\lambda _{0}$. Introducing
the angular variable $\theta $ in accordance with $\lambda _{0}=(m\omega
_{0}/c)\sqrt{\varepsilon _{0}}\sin \theta $, $0\leqslant \theta \leqslant
\pi $, the radiation intensity is written in the form%
\begin{equation}
I_{(1)}\approx \frac{2q^{2}\beta ^{2}}{c\sqrt{\varepsilon _{0}}}%
\sum_{m=1}^{\infty }m^{2}\omega _{0}^{2}\int_{0}^{\pi /2}d\theta \,\sin
\theta J_{m}^{\prime 2}(\beta \sin \theta ),  \label{Ilim2}
\end{equation}%
where $\beta =v\sqrt{\varepsilon _{0}}/c$.

In the limit $\rho _{1}\rightarrow \infty $, for the second type of the
modes one has%
\begin{equation}
\varepsilon _{0}|\lambda _{1}|\frac{J_{m}^{\prime }}{J_{m}}+\varepsilon
_{1}\lambda _{0}\frac{K_{m}^{\prime }}{K_{m}}\approx -\varepsilon
_{0}|\lambda _{1}|\tan \left( \lambda _{0}\rho _{1}-\frac{m}{2}\pi -\frac{%
\pi }{4}\right) -\varepsilon _{1}\lambda _{0}=0.  \label{limMode2}
\end{equation}%
In the way similar to the case of the modes (\ref{limMode1}), the
corresponding contribution to the radiation intensity can be presented in
the form%
\begin{equation}
I_{(2)}\approx \frac{2q^{2}\omega _{0}^{2}}{c\sqrt{\varepsilon _{0}}}%
\sum_{m=1}^{\infty }m^{2}\int_{0}^{\pi /2}d\theta \frac{J_{m}^{2}(m\beta
\sin \theta )}{\sin \theta }\cos ^{2}\theta .  \label{Ilim3}
\end{equation}%
Summing the contribution from the modes of the first and second types, given
by (\ref{Ilim2}) and (\ref{Ilim3}), we find the intensity for the
synchrotron radiation in a homogeneous medium with dielectric permittivity $%
\varepsilon _{0}$ (for the properties of the synchrotron radiation in a
homogeneous medium see \cite{Tsyt51,Kita60,Zrel70}).

Alternatively, the radiation intensity inside the dielectric cylinder can be
obtained by evaluating the energy flux through the cross-section of the
dielectric cylinder:%
\begin{equation}
I_{f}^{\text{(in)}}=\int_{0}^{\rho _{1}}d\rho \int_{0}^{2\pi }d\phi \,\rho
\mathbf{n}_{z}\cdot \mathbf{S}=\frac{c}{4\pi }\int_{0}^{\rho _{1}}d\rho
\int_{0}^{2\pi }d\phi \,\rho \left( E_{\rho }H_{\phi }-E_{\phi }H_{\rho
}\right) ,  \label{If}
\end{equation}%
where $\mathbf{n}_{z}$ is the unit vector along the $z$-axis and $\mathbf{S}%
=c\left[ \mathbf{E}\times \mathbf{H}\right] /(4\pi )$ is the Poynting
vector. By using the expressions for the corresponding components of the
magnetic and electric fields from (\ref{Hrad}) and (\ref{Erad}), we find the
expression below:
\begin{eqnarray}
I_{f}^{\text{(in)}} &=&\sum_{m=1}^{\infty }I_{f,m}^{\text{(in)}}=\frac{%
q^{2}v^{2}\rho _{1}^{2}}{8\varepsilon _{0}\omega _{0}}\sum_{m=1}^{\infty
}\sum_{s}\frac{k_{m,s}\lambda _{m,s}^{(1)2}J_{m}^{-4}}{m\alpha _{m}^{\prime
2}(k_{m,s})K_{m}^{2}}\left[ \sum_{l=\pm 1}\frac{J_{m+l}(\lambda _{m,s}\rho
_{0}/\rho _{1})}{V_{m}-lmu}\right] ^{2}  \notag \\
&&\times \sum_{p=\pm 1}\frac{K_{m+p}}{V_{m}-pmu}\left[ \left( \frac{%
m^{2}\omega _{0}^{2}}{c^{2}}\varepsilon _{0}+k_{m,s}^{2}\right) \frac{K_{m+p}%
}{V_{m}-pmu}-\frac{\lambda _{m,s}^{2}}{\rho _{1}^{2}}\frac{K_{m-p}}{V_{m}+pmu%
}\right]   \notag \\
&&\times \left[ J_{m+p}^{\prime 2}+\left( 1-\frac{(m+p)^{2}}{\lambda
_{m,s}^{2}}\right) J_{m+p}^{2}\right] .  \label{If1}
\end{eqnarray}%
Note that $I_{f}^{\text{(in)}}$ is the flux in the region $z>0$. By the
symmetry of the problem we have the same flux in the region $z<0$ and the
total flux will be $2I_{f}^{\text{(in)}}$. In a way similar to that used for
formula (\ref{Iwork2}), it can be seen that in the limit $\rho
_{1}\rightarrow \infty $, from (\ref{If1}) the expression for the radiation
intensity in a homogeneous medium is obtained.

\begin{figure}[tbph]
\begin{center}
\epsfig{figure=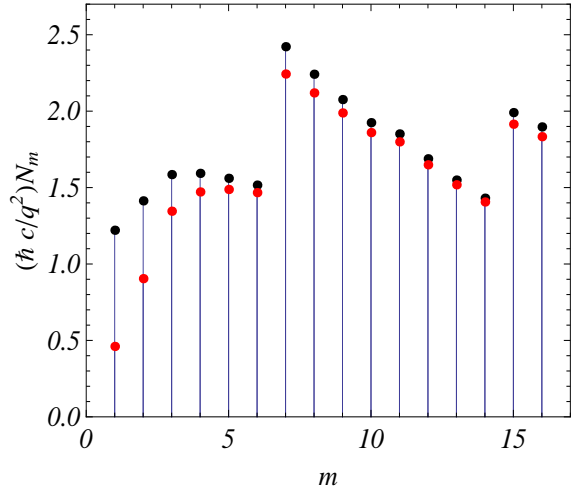, width=7.5cm, height=6.5cm}
\end{center}
\caption{The number of the radiated quanta per period of the rotation on a
given mode $m$ for the electron energy $E_{e}=2$ MeV and for the values of
the parameters $\protect\varepsilon _{1}=1$, $\protect\varepsilon _{0}=3$, $%
\protect\rho _{1}/\protect\rho _{0}=1.2$. The black/red points correspond to
the number of quanta evaluated by formula (\protect\ref{Nm})/(\protect\ref%
{Nmf}).}
\label{fig1}
\end{figure}

In figure \ref{fig1}, by the black points we present the number of the
radiated quanta at a given harmonic $m$ per period of the charge rotation,%
\begin{equation}
N_{m}=\sum_{s}N_{m,s}=\frac{TI_{m}}{\hbar m\omega _{0}},\;T=\frac{2\pi }{%
\omega _{0}},  \label{Nm}
\end{equation}%
for the electron energy $E_{e}=2$ MeV and for the values of the parameters $%
\varepsilon _{1}=1$, $\varepsilon _{0}=3$, $\rho _{1}/\rho _{0}=1.2$. The
red points correspond to the number of the radiated quanta evaluated by the
flux through the cylinder cross-section:%
\begin{equation}
N_{f,m}^{\text{(in)}}=\frac{2TI_{f,m}^{\text{(in)}}}{\hbar m\omega _{0}}.
\label{Nmf}
\end{equation}%
The factor 2 in the last expression corresponds to that we have the same
amount of the energy flux in the region $z<0$. As we have already noted
above, for the number of the modes one has $s_{m}=1$ for $m\leqslant 6$, $%
s_{m}=2$ for $7\leqslant m\leqslant 10$, $s_{m}=3$ for $11\leqslant
m\leqslant 14$, and $s_{m}=4$ for $m=15,16$. As it is seen from the graph
the number of the radiated quanta increases with the appearance of new
modes. The numerical data in figure \ref{fig1} show that $N_{m}\neq N_{f,m}^{%
\text{(in)}}$, i.e., the number of the radiated quanta evaluated by the
energy flux through the cross-section of the cylinder and by the work done
by the radiation field differ. As it will be shown in the next section, the
reason for this difference is that for the eigenmodes of the dielectric
cylinder there is also energy flux in the exterior region located near the
surface of the cylinder.

\begin{figure}[tbph]
\begin{center}
\epsfig{figure=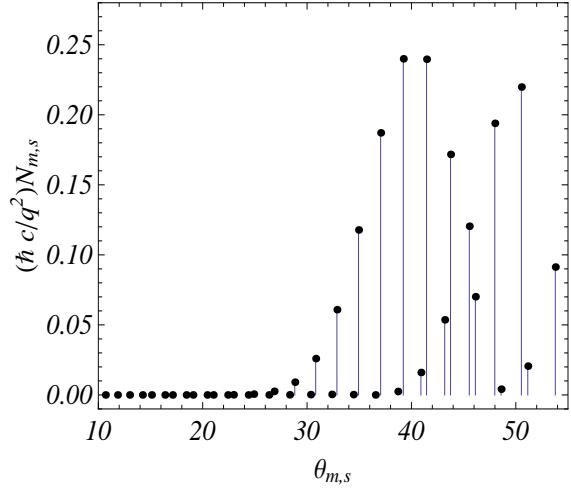, width=7.5cm, height=6.5cm}
\end{center}
\caption{The number of the radiated quanta per period of the rotation on a
given mode as a function of $\protect\theta _{m,s}$. The values of the
parameters are as follows: $E_{e}=2$ MeV, $\protect\varepsilon _{1}=1$, $%
\protect\varepsilon _{0}=3$, $m=16$ and $\protect\rho _{1}/\protect\rho %
_{0}=4$.}
\label{fig2}
\end{figure}

As it has been mentioned above, the number of the modes for a given $m$
increases with increasing $\rho _{1}$. For example, in the case $m=16$ one
has $s_{m}=14$ for $\rho _{1}/\rho _{0}=2$ and $s_{m}=40$ for $\rho
_{1}/\rho _{0}=4$ with the values of the other parameters being the same as
those for figure \ref{fig1}. In order to illustrate the dependence on $s$,
in figure \ref{fig2} we present the number of the radiated quanta at a given
mode, defined by (\ref{Nm}), as a function of $\theta _{m,s}$ for $E_{e}=2$
MeV, $\varepsilon _{1}=1$, $\varepsilon _{0}=3$, $m=16$ and $\rho _{1}/\rho
_{0}=4$. For large values $s$, the separation of two types of the modes is
seen which are interlaced.

\section{Radiation intensity outside a cylinder}

\label{sec:IntExt}

For a charge rotating inside a dielectric cylinder, in Ref. \cite{Kota02a}
we have evaluated the average energy flux per unit time through the
cylindrical surface of large radius $\rho $ ($\rho \gg \rho _{1}$) coaxial
with the cylinder:%
\begin{equation}
I_{\text{ex}}=\frac{2\pi }{T}\int_{0}^{T}dt\,\int_{-\infty }^{+\infty
}dz\,\rho \mathbf{n}_{\rho }\cdot \mathbf{S}=\sum_{m=1}^{\infty
}\int_{0}^{\pi }d\theta \,\frac{dI_{m}^{\text{(ex)}}}{d\theta },
\label{Iext}
\end{equation}%
where $\mathbf{n}_{\rho }$ is the unit vector along the radial direction and
$\theta $ is the angle between the radiation direction and the axis of the
cylinder. The expression for the angular density of the radiation intensity
on a given harmonic $m$ with the frequency $m\omega _{0}$ is given by the
expression
\begin{equation}
\frac{dI_{m}^{\text{(ex)}}}{d\theta }=\frac{q^{2}v^{2}m^{2}\omega _{0}^{2}%
\sqrt{\varepsilon _{1}}}{\pi ^{2}c^{3}\rho _{1}^{2}}\sin \theta \left[
|C_{m}^{(1)}-C_{m}^{(-1)}|^{2}+|C_{m}^{(1)}+C_{m}^{(-1)}|^{2}\cos ^{2}\theta %
\right] ,  \label{dIext}
\end{equation}%
where we have defined%
\begin{equation}
C_{m}^{(p)}=\frac{J_{m+p}(\lambda _{0}^{\text{(ex)}}\rho _{0})}{W_{m+p}^{J}}%
+p\lambda _{1}^{\text{(ex)}}H_{m}(\lambda _{1}^{\text{(ex)}}\rho _{1})\frac{%
J_{m+p}(\lambda _{0}^{\text{(ex)}}\rho _{1})}{2\alpha _{m}W_{m+p}^{J}}%
\sum_{l=\pm 1}\frac{J_{m+l}(\lambda _{0}^{\text{(ex)}}\rho _{0})}{W_{m+l}^{J}%
},  \label{Cmp}
\end{equation}%
and%
\begin{equation}
\lambda _{0}^{\text{(ex)}}=\frac{m\omega _{0}}{c}\sqrt{\varepsilon
_{0}-\varepsilon _{1}\cos ^{2}\theta },\;\lambda _{1}^{\text{(ex)}}=\frac{%
m\omega _{0}}{c}\sqrt{\varepsilon _{1}}\sin \theta .  \label{lam01tet}
\end{equation}

The properties of the radiation corresponding to (\ref{dIext}) are described
in detail in \cite{Kota02a} (see also Ref. \cite{Saha05} for a more general
case of helical motion inside a dielectric cylinder). In particular, it was
shown that under the Cherenkov condition for dielectric permittivity of the
cylinder and the velocity of the particle image on the cylinder surface,
strong narrow peaks are present in the angular distribution for the number
of radiated quanta. At these peaks the radiated energy exceeds the
corresponding quantity for a homogeneous medium by several orders of
magnitude. The angular locations of the peaks are determined from the
equation which is obtained from the equation determining the eigenmodes for
the dielectric cylinder by the replacement $H_{m}(\lambda _{1}^{\text{(ex)}%
}\rho _{1})\rightarrow Y_{m}(\lambda _{1}^{\text{(ex)}}\rho _{1})$, where $%
Y_{m}(x)$ is the Neumann function. For the illustration, in figure \ref{fig3}
we plot the angular density of the number of the quanta radiated in the
exterior region per period of the charge rotation,%
\begin{equation}
\frac{dN_{m}^{\text{(ex)}}}{d\theta }=\frac{T}{\hbar m\omega _{0}}\frac{%
dI_{m}^{\text{(ex)}}}{d\theta },  \label{Nmext}
\end{equation}%
with the radiation intensity given by (\ref{dIext}). The corresponding
parameters are as follows: $E_{e}=2$ MeV, $\varepsilon _{1}=1$, $\varepsilon
_{0}=3$, $m=16$, $\rho _{1}/\rho _{0}=1.2$. The dashed curve corresponds to
the synchrotron radiation in the vacuum in the absence of the dielectric
cylinder ($\varepsilon _{0}=1$). We see the presence of the strong narrow
peak at $\theta \approx 42.63$ (in degrees). The height of this peak is $%
\approx 463.4$ and the width $\approx 0.05$. At large distances from the
cylinder for the total number of the radiated quanta one has $N_{m}^{\text{%
(ex)}}=1.748$. Note that for the number of quanta radiated on the eigenmodes
of the cylinder we have $N_{m}=1.897$ (see figure \ref{fig1}). For the
synchrotron radiation from an electron in the vacuum ($\varepsilon _{0}=1$)
one has $N_{m}^{(0)}=0.352$. As it is seen, the presence of the cylinder
considerably increases the radiation intensity.
\begin{figure}[tbph]
\begin{center}
\epsfig{figure=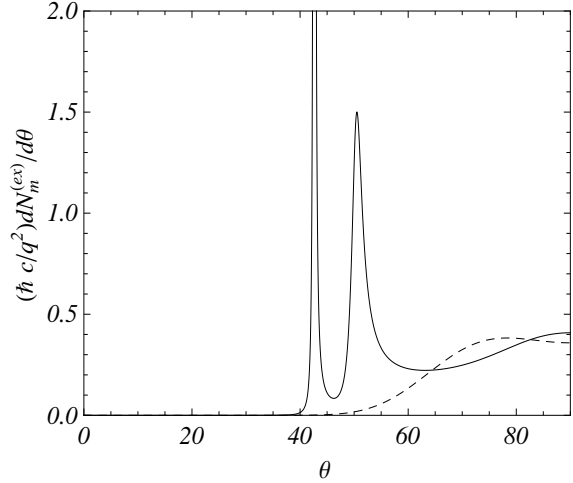, width=7.5cm, height=6.5cm}
\end{center}
\caption{The angular density of the number of the quanta radiated on the
harmonic $m=16$ in the exterior region per period of the charge rotation as
a function of $\protect\theta $ (in degrees). The full curve is plotted for
the values of parameters $E_{e}=2$ MeV, $\protect\varepsilon _{1}=1$, $%
\protect\varepsilon _{0}=3$, $m=16$, $\protect\rho _{1}/\protect\rho _{0}=1.2
$. The dashed curve is for the radiation in the vacuum ($\protect\varepsilon %
_{0}=1$).}
\end{figure}

Now we shall show that, in addition to the radiation corresponding to (\ref%
{Iext}), in the exterior region we have also radiation fields localized near
the surface of the dielectric cylinder. These fields are the tails of the
eigenmodes for the dielectric cylinder. For this we need the expressions of
the electric and magnetic fields in the exterior region. These fields can be
presented in the form of the Fourier expansion (\ref{FlFour}), with the
Fourier components of the magnetic field given by the expressions \cite%
{Kota02a}
\begin{eqnarray}
H_{ml} &=&\frac{i^{2-\sigma _{l}}qvk_{z}}{2\pi c}\sum_{p=\pm 1}p^{\sigma
_{l}-1}B_{2,m}^{(p)}H_{m+p}(\lambda _{1}\rho ),\;l=\rho ,\phi ,  \notag \\
H_{mz} &=&-\frac{qv\lambda _{1}}{2\pi c}\sum_{p=\pm
1}pB_{2,m}^{(p)}H_{m}(\lambda _{1}\rho ),  \label{Hmext}
\end{eqnarray}%
where
\begin{equation}
B_{2,m}^{(p)}=\frac{J_{m+p}(\lambda _{0}\rho _{0})}{\rho _{1}W_{m+p}^{J}}+%
\frac{p\lambda _{1}J_{m+p}(\lambda _{0}\rho _{1})}{2\rho _{1}\alpha
_{m}W_{m+p}^{J}}H_{m}(\lambda _{1}\rho _{1})\sum_{l=\pm 1}\frac{%
J_{m+l}(\lambda _{0}\rho _{0})}{W_{m+l}^{J}},  \label{B2m}
\end{equation}%
and the other notations are the same as in (\ref{Hmz}). For the electric
field one has the expressions%
\begin{eqnarray}
E_{ml} &=&\frac{qvi^{1-\sigma _{l}}}{4\pi m\omega _{0}\varepsilon _{1}}%
\sum_{p=\pm 1}p^{\sigma _{l}}H_{m+p}(\lambda _{1}\rho )\sum_{j=\pm 1}j\left(
\frac{m^{2}\omega _{0}^{2}}{c^{2}}\varepsilon _{1}+jk_{z}^{2}\right)
B_{2,m}^{(jp)},  \notag \\
E_{mz} &=&-\frac{qiv\lambda _{1}k_{z}}{2\pi m\omega _{0}\varepsilon _{1}}%
\sum_{p=\pm 1}B_{2,m}^{(p)}H_{m}(\lambda _{1}\rho ).  \label{Emext}
\end{eqnarray}

For a fixed value of $\rho >\rho _{1}$ and in the limit $z\rightarrow \infty
$ the radiation fields are determined by the poles of the integrand in (\ref%
{FlFour}). These poles correspond to the zeros of the function $\alpha _{m}$%
. The fields are evaluated in a way similar to that we have used for the
interior region. They are presented in the form (\ref{Flr}), where
\begin{eqnarray}
H_{m,s}^{l}(\rho ) &=&k_{m,s}\sum_{p=\pm 1}p^{\sigma
_{l}-1}B_{2,m,s}^{(p)}K_{m+p}(\lambda _{m,s}^{(1)}\rho /\rho _{1}),  \notag
\\
H_{m,s}^{z}(\rho ) &=&\frac{\lambda _{m,s}}{\rho _{1}}\sum_{p=\pm
1}B_{2,m,s}^{(p)}K_{m}(\lambda _{m,s}^{(1)}\rho /\rho _{1}),
\label{Hmextrad}
\end{eqnarray}%
for the magnetic field and
\begin{eqnarray}
E_{m,s}^{l}(\rho ) &=&\frac{(-1)^{\sigma _{l}-1}c}{2\varepsilon _{1}m\omega
_{0}}\sum_{p=\pm 1}p^{\sigma _{l}}K_{m+p}(\lambda _{m,s}^{(1)}\rho /\rho
_{1})\sum_{j=\pm 1}\left( \frac{m^{2}\omega _{0}^{2}}{c^{2}}\varepsilon
_{1}+jk_{m,s}^{2}\right) B_{2,m,s}^{(jp)},  \notag \\
E_{m,s}^{z}(\rho ) &=&\frac{c\lambda _{m,s}k_{m,s}}{\varepsilon _{1}\rho
_{1}m\omega _{0}}\sum_{p=\pm 1}pB_{2,m,s}^{(p)}K_{m}(\lambda
_{m,s}^{(1)}\rho /\rho _{1}),  \label{Emextrad}
\end{eqnarray}%
for the electric field. In these expressions we have defined%
\begin{equation}
B_{2,m,s}^{(p)}=\frac{\lambda _{m,s}^{(1)}J_{m+p}/K_{m}}{\left(
V_{m}-pmu\right) J_{m}^{2}}\sum_{l=\pm 1}\frac{J_{m+l}(\lambda _{m,s}\rho
_{0}/\rho _{1})}{V_{m}-lmu}.  \label{B2ms}
\end{equation}%
As it is seen from (\ref{Hmextrad}) and (\ref{Emextrad}), the radiation
fields corresponding to the eigenmodes of the dielectric cylinder
exponentially decay in the exterior region for $\rho /\rho _{1}\gg 1/\lambda
_{m,s}^{(1)}$.

For the corresponding energy flux through a plane perpendicular to the axis
of the cylinder,%
\begin{equation}
I_{f}^{\text{(ex)}}=\int\limits_{\rho _{1}}^{\infty }d\rho \,\rho
\int\limits_{0}^{2\pi }d\phi \,\mathbf{n}_{z}\cdot \mathbf{S},  \label{Ifex}
\end{equation}%
one finds%
\begin{eqnarray}
I_{f}^{\text{(ex)}} &=&\sum_{m=1}^{\infty }I_{f,m}^{\text{(ex)}}=\frac{%
q^{2}v^{2}\rho _{1}^{2}}{8\varepsilon _{1}\omega _{0}}\sum_{m=1}^{\infty
}\sum_{s}\frac{k_{m,s}\lambda _{m,s}^{(1)2}J_{m}^{-4}}{m\alpha _{m}^{\prime
2}(k_{m,s})K_{m}^{2}}\left[ \sum_{l=\pm 1}\frac{J_{m+l}(\lambda _{m,s}\rho
_{0}/\rho _{1})}{V_{m}-lmu}\right] ^{2}  \notag \\
&&\times \sum_{p=\pm 1}\frac{J_{m+p}}{V_{m}-pmu}\left[ \left( \frac{%
m^{2}\omega _{0}^{2}}{c^{2}}\varepsilon _{1}+k_{m,s}^{2}\right) \frac{J_{m+p}%
}{V_{m}-pmu}-\frac{\lambda _{m,s}^{(1)2}}{\rho _{1}^{2}}\frac{J_{m-p}}{%
V_{m}+pmu}\right]  \notag \\
&&\times \left[ K_{m+p}^{\prime 2}-\left( 1+\frac{(m+p)^{2}}{\lambda
_{m,s}^{(1)2}}\right) K_{m+p}^{2}\right] .  \label{Ifexm}
\end{eqnarray}%
Now it can be explicitly checked that one has the relation $I_{m}=2\left(
I_{f,m}^{\text{(in)}}+I_{f,m}^{\text{(ex)}}\right) $.

\begin{figure}[tbph]
\begin{center}
\epsfig{figure=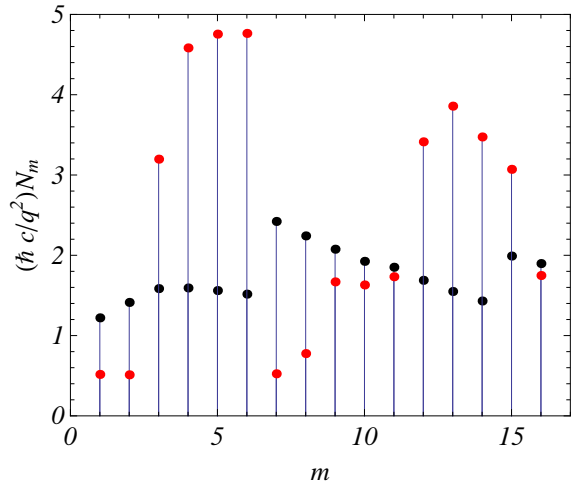, width=7.5cm, height=6.5cm}
\end{center}
\caption{The number of the radiated quanta per period of the rotation on a
given mode $m$ for different values of $m$ and for the values of the
parameters $E_{e}=2$ MeV, $\protect\varepsilon _{1}=1$, $\protect\varepsilon %
_{0}=3$, $\protect\rho _{1}/\protect\rho _{0}=1.2$. The black points
correspond to the radiation in the exterior region and the red points are
for the radiation emitted on the eigenmodes of the dielectric cylinder.}
\label{fig4}
\end{figure}

\begin{figure}[tbph]
\begin{center}
\epsfig{figure=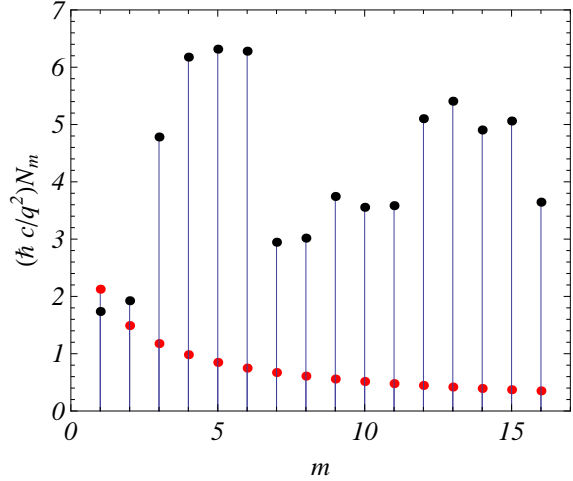, width=7.5cm, height=6.5cm}
\end{center}
\caption{The total number of the radiated quanta, $N_{m}=N_{m}^{\text{(in)}%
}+N_{m}^{\text{(ex)}}$, in the presence (black points) and in the absence
(red points) of the dielectric cylinder. The values of the parameters are
the same as those for figure \protect\ref{fig4}.}
\label{fig5}
\end{figure}

In figure \ref{fig4} we have plotted the number of the radiated quanta per
period of the rotation on a given mode $m$ for different values of $m$. The
values of the parameters are as follows: $E_{e}=2$ MeV, $\varepsilon _{1}=1$%
, $\varepsilon _{0}=3$, $\rho _{1}/\rho _{0}=1.2$. The black points
correspond to the radiation in the exterior region at latge distances from
the cylinder, $N_{m}^{\text{(ex)}}=\int_{0}^{\pi /2}d\theta (dN_{m}^{\text{%
(ex)}}/d\theta )$ (see ()\ref{Nmext}), and the red points are for the
radiation on the eigenmodes of the dielectric cylinder, $N_{m}^{\text{(in)}%
}=\sum_{s}N_{m,s}$ (see (\ref{Nm})). For the comparison with the synchrotron
radiation in the vacuum, in figure \ref{fig5}, for the same values of the
parameters, we plot the total number of the radiated quanta, $N_{m}=N_{m}^{%
\text{(in)}}+N_{m}^{\text{(ex)}}$, in the presence (black points) and in the
absence (red points) of the dielectric cylinder. As it is seen from these
graphs, the presence of the cylinder essentially increases the radiation
intensity.

We have considered the energy loss for a particle rotating in a medium due
to the synchrotron radiation. A point which deserves a separate
investigation is the role of the other processes of the particle interaction
with the medium. In particular, they include ionization energy losses,
particle bremsstrahlung in media, and the multiple scattering (see, for
instance, \cite{TerMik72,Akhi96,Rull98}). The relative role of these
processes depends on the particle energy and characteristics of the medium
and has been shortly discussed in \cite{Saha07b}. An interesting possibility
of escaping ionization losses in the medium was indicated in \cite{Bolo61},
where it was argued that a narrow empty channel along the particle
trajectory in the solid dielectric does not affect the radiation intensity
if the channel radius is less than the radiation wavelength.

\section{Conclusion}

\label{sec:Conc}

We have investigated the synchrotron radiation from a charged particle
rotating inside a dielectric cylinder surrounded by a homogeneous medium.
The radiation intensity in the exterior medium at large distances from the
cylinder has been considered previously in \cite{Kota02a} and here we were
mainly concerned with the radiation propagating inside the cylinder. The
electromagnetic fields inside the cylinder are presented in the form of the
Fourier expansion (\ref{vecpot21}), where the Fourier components for the
magnetic and electric fields are given by the expressions (\ref{Hmz}) and (%
\ref{Emlz}). The radiation parts of the fields propagating inside the
cylinder are determined by the contribution of the poles of the Fourier
components. These poles correspond to the zeros of the function $\alpha _{m}$
given by (\ref{bet1}). The latter are the bound modes of the dielectric
cylinder and are determined from the equation (\ref{eigmodesnew2}). The
electromagnetic fields for the radiation propagating inside the cylinder are
presented in the form (\ref{Flr}) with the expressions (\ref{Hrad}) and (\ref%
{Erad}) for separate components of the magnetic and electric fields.

The radiation intensity inside the cylinder is investigated in Section \ref%
{sec:IntensInt}. Firstly we have evaluated the work done by the radiation
field on the charged particle. The radiation intensity evaluated in this way
is given by the expression (\ref{Iwork2}). In the limit when the radius of
the cylinder goes to infinity this expression reduces to the formula for the
intensity of the synchrotron radiation in a homogeneous medium with
dielectric permittivity $\varepsilon _{0}$. Further, we have evaluated the
energy flux through the cross-section of the cylinder. The radiation
intensity evaluated in this way is given by the expression (\ref{If1}). A
numerical example for the radiation intensities evaluated in two different
ways is presented in figure \ref{fig1}. As it is seen, the intensities
evaluated by the work done on the charge and by the flux through the
cross-section do not coincide. In Section \ref{sec:IntExt} we show that the
reason for this difference is that for the eigenmodes of the dielectric
cylinder there is also energy flux in the exterior region located near the
surface of the cylinder.

The radiation field in the exterior region consists of two parts. The first
one corresponds to the radiation propagating at large distances from the
cylinder. The angular density of the corresponding radiation intensity on a
given harmonic $m$ is given by the formula (\ref{dIext}). For large values
of $m$ and under the Cherenkov condition for dielectric permittivity of the
cylinder and the velocity of the particle image on the cylinder surface,
strong narrow peaks appear in the angular distribution for the radiation
intensity. An example is given in figure \ref{fig3}. The second part of the
radiation field in the exterior region corresponds to the modes of the
dielectric cylinder. The corresponding fields are located near the surface
of the cylinder and are given by the formula (\ref{FlFour}), with the
Fourier components of the magnetic and electric fields given by the
expressions (\ref{Hmext}) and (\ref{Emext}). These fields exponentially
decay at distances from the cylinder surface larger than the radiation
wavelength. The energy flux in the exterior region corresponding to the
eigenmodes of the cylinder is given by the expression (\ref{Ifexm}). We have
explicitly checked that the radiation intensity evaluated by the work done
by the radiation field on the charged particle is equal to the radiation
intensity evaluated by the energy flux through the plane perpendicular to
the cylinder axis if we take into account the contribution of the radiation
field in the exterior region.

As we have seen in the present paper, the insertion of a dielectric
waveguide provides an additional mechanism for tuning the characteristics of
the synchrotron radiation by choosing the parameters of the waveguide. The
radiated energy inside the cylinder is redistributed among the cylinder
modes and, as a result, the corresponding spectrum differs significantly
from the homogeneous medium or free-space results. This change is of special
interest in the low-frequency range where the distribution of the radiation
energy among small number of modes leads to the enhancement of the spectral
density for the radiation intensity. The radiation emitted on the waveguide
modes propagates inside the cylinder and the waveguide serves as a natural
collector for the radiation. This eliminates the necessity for focusing to
achieve a high-power spectral intensity. The geometry considered here is of
interest also from the point of view of generation and transmitting of waves
in waveguides, a subject which is of considerable practical importance in
microwave engineering and optical fiber communications.

\section*{Acknowledgement}

The authors are grateful to Professor L. Sh. Grigoryan, S. R. Arzumanyan, H.
F. Khachatryan for stimulating discussions.

\end{document}